\pdfoutput=1
\documentclass{article}
\usepackage[T1]{fontenc} 
\usepackage[utf8]{inputenc} 
\usepackage{ismir,amsmath,cite,url}
\usepackage{graphicx}
\usepackage{color}
\usepackage{textgreek}
\usepackage[neverdecrease]{paralist}

\usepackage{lineno}

\title{Collaborative Song Dataset (CoSoD): An annotated dataset of multi-artist collaborations in popular music}

\multauthor
{Michèle Duguay$^1$ \hspace{1cm} Kate Mancey$^1$ \hspace{1cm} Johanna Devaney$^2$} { \\
 $^1$ Department of Music, Harvard University\\
$^2$ Brooklyn College and Graduate Center, City University of New York\\
{\tt\small mduguay@fas.harvard.edu, johanna.devaney@brooklyn.cuny.edu}
}

\def\authorname{M. Duguay, K. Mancey, and J. Devaney}

\begin{document}

\maketitle
\begin{abstract}
The Collaborative Song Dataset (CoSoD) is a corpus of 331 multi-artist collaborations from the 2010–2019 \textit{Billboard} “Hot 100” year-end charts. The corpus is annotated with formal sections, aspects of vocal production (including reverberation, layering, panning, and gender of the performers), and relevant metadata. CoSoD complements other popular music datasets by focusing exclusively on musical collaborations between independent acts. In addition to facilitating the study of song form and vocal production, CoSoD allows for the in-depth study of gender as it relates to various timbral, pitch, and formal parameters in musical collaborations. In this paper, we detail the contents of the dataset and outline the annotation process. We also present an experiment using CoSoD that examines how the use of reverberation, layering, and panning are related to the gender of the artist. In this experiment, we find that men's voices are on average treated with less reverberation and occupy a more narrow position in the stereo mix than women's voices.
\end{abstract}
\section{Introduction}\label{sec:introduction}

As far back as the 1960s, \textit{Billboard} charts have featured collaborations between independent acts. In recent years, however, the number of songs featuring a collaboration between artists has skyrocketed \cite{Economist2018}. Part of this is due to the rising popularity of hip-hop in the 1980s, in which collaboration between different artists is a fixture. The 1986 version of “Walk This Way” by Aerosmith and Run DMC is an oft-cited example of such a collaboration. As Rose notes, the success of a collaboration between a hip-hop group (Run DMC) and a rock group (Aerosmith) “brought [hip-hop’s] strategies of intertextuality into the commercial spotlight” \cite[p. 51--52]{Rose1994}. The 1990 success of “She Ain’t Worth It” by Glenn Medeiros ft. Bobby Brown marked the first time a sung and rapped collaboration reached \#1 on \textit{Billboard}’s “Hot 100.” Molanphy notes that during this period, multi-artist collaborations crystallized into two different frameworks: the “featured bridge rapper,” and the “featured hook singer” \cite{Molanphy2015}. Subsequently, tracks with one or more guest artist(s) have become a mainstay on the charts. 


By 2021, over a third (39\%) of the songs in \textit{Billboard}'s “Hot 100” year-end chart credited more than one artist. Consider for instance “Save Your Tears,” by singers The Weeknd \& Ariana Grande, which occupied second place on the chart. A solo version of the song originally appeared on The Weeknd’s album \textit{After Hours} (2020). While this version achieved commercial success, the remix with Ariana Grande became a \#1 single on the \textit{Billboard} “Top 100” in May 2021 and became the longest-charting collaboration in \textit{Billboard} “Hot 100” history. In the remix, Grande performs approximately half of the vocals, transforming the solo song into a dialogue between two characters. The collaboration between the two artists is responsible for the popularity of the remix, inviting both Grande’s and The Weeknd’s fans to stream, buy, and otherwise engage with the song. Several musicological studies have examined this relationship between collaborative songs and commercial success \cite{Ordanini2018, Silva2019, Oliveira2020}. Other work has provided in-depth explorations of the musical characteristics of collaborative songs, with a particular focus on hip-hop \cite{Komaniecki2017, Duinker2020, Duguay2022}.



Given the popularity of multi-artist collaborations, a more systematic exploration of their musical features is warranted. In this paper, we introduce the Collaborative Song Dataset (CoSoD), an annotated dataset that facilitates the study of various musical features in multi-artist collaborations. CoSoD provides metadata and analytical data for 331 multi-artist collaborations appearing on the \textit{Billboard} “Hot 100” year-end charts between 2010 and 2019. The dataset also provides timed annotations on the song's formal structure, artists' gender, vocal delivery and pitch, and vocal production (reverberation, panning, and layering). As detailed in Section 2, the range of features included in the dataset makes it more broadly applicable for MIR research tasks. These include structural segmentation, vocal mixing, automatic music production, and examinations of gender in popular music. After outlining the contents of the dataset and the annotation methodology in Section 3, we present an experiment in Section 4 that examines the relationship between vocal production parameters and the gender of the performer in a subset of CoSoD. 

\section{Related Work}\label{sec:releated}

CoSoD complements the growing list of annotated datasets that provide information on song structure in various popular music genres, e.g.,\cite{Berenzweig2003, Harte2010, Burgoyne2011, Smith2011, Bertin-Mahieux2011, deClerq2011, Bimbot2014, Nieto2019}, and is the first dataset to exclusively contain data on collaborative songs between independent acts. It can thus be used for training and evaluating structural segmentation tasks and for studying the specific structural characteristics of collaborative songs. CoSoD also complements existing datasets for multi-track mixing/analysis\cite{Hsu2010, Bittner2014, Chan2015, bittner2016, Man2017, Rafii2017} and vocal analysis\cite{gong2017, Wang2018, wilkins2018} by providing analytical annotations on the treatment of the voice in a mix. 

In recent years, several studies have proposed tools and methods to automate the mixing of multi-track recordings\cite{DeMan2017, martínezramírez2022}. Such automatic production methods have various artistic and creative applications. One framework has been suggested to remix early jazz recordings, which are pre-processed using source separation then remixed with automatic production tools\cite{Matz2015}. \cite{Ishizaki2009} proposes a prototype for an automatic DJ mixing system allowing for cross-fading via beat and tempo adjustment between songs. Studies on automatic mixing can be enhanced by knowledge of common mixing practices for specific instruments or sound sources. For instance, one study uses mixing practices that are consistent between mixing engineers to create a model that automatically mixes multiple drum tracks\cite{Scott2013}. By focusing on vocals, which are a salient component of the mix in popular music\cite{Knoll2022}, CoSoD provides a complementary approach to these studies on automated production. By providing annotations based on close listening of specific vocal mixing parameters in the different formal sections of a song, the dataset allows for the identification of trends in panning, layering, and use of artificial reverberation as they are applied to vocals in commercially successful post-2010 popular music. It enables the direct comparison of how various mixing parameters are applied to individual artists' voices within and across songs. In addition to facilitating the modeling of voice mixing, CoSoD also allows musicologists to ask questions about the way different voice types and individuals are mixed. 

Finally, CoSoD facilitates the study of the relationship between gender and popular music. A number of previous studies have examined music programming and streaming services, exploring for instance how listeners tend to stream male artists more than women and mixed-gender groups\cite{EppsDarling2020}. Watson discusses gender inequality and low programming of women’s music in country music radio\cite{Watson2020}. Other work addresses how a listener’s declared gender impacts automatic music recommendation\cite{Vigliensoni2016} and musical preferences\cite{Laplante2014}. Additionally, various studies have addressed race and gender, along with sexist and racist discourses and practices, as they impact the music industry in general and the \textit{Billboard} charts in particular\cite{Keyes2000,Dowd2002,Bradley2015,Lafrance2018,Lieb2018,Watson2019,bauer2022}. By providing data on musical features, gender, and the role of these parameters within the formal structure of a song, CoSoD offers a new and complementary angle for the study of gender as it directly relates to the musical content of post-2010 popular collaborations. 

\section{Collaborative Song Dataset (CoSoD)}\label{sec:cosod}

CoSoD\footnote{\url{https://github.com/duguay-michele/CoSoD}} consists of metadata and analytical data of a 331-song corpus comprising all multi-artist collaborations on the \textit{Billboard} “Hot 100” year-end charts published between 2010 and 2019. Each song in the dataset is associated with two CSV files: one for metadata and one for analytical data. We assembled the corpus by identifying every song on the charts that featured collaborations between two or more artists who usually perform independently from one another. 

\subsection{Annotation of Musical Features}
The following analytical data is provided for each song in the dataset:

\setdefaultleftmargin{0.4cm}{}{}{}{}{}
\begin{enumerate}
    \setlength\itemsep{0.2em}
    \item \textbf{Index number:} 1 to 33
    \item \textbf{Time stamps:} In seconds (start of new  section)
    \item \textbf{Formal section label:} \textit{Introduction, Verse, Pre-chorus, Chorus, Hook, Dance Chorus\cite{Barna2020}, Link, Post-chorus, Bridge, Outro, Refrain} or \textit{Other}
    \item \textbf{Name of artist(s):} Full name of the artist performing in each section. If all artists credited on the \textit{Billboard} listing perform in a section, the label \textit{both} or \textit{all} is used.
\end{enumerate}

Songs were assigned at random to one of two annotators, who generated time stamps at the onset of each formal section with Sonic Visualiser.\footnote{The first annotator (first author) has a doctorate in music theory, while the second (second author) is a doctoral candidate in the same field.} The annotators provided formal labels according to their analysis of the song. In case of ambiguity in the formal sections, both annotators discussed the analysis and agreed upon an interpretation.  

For each formal section performed by \textit{one artist only}, the following analytical data on the voice is provided:

\begin{enumerate}
 \setlength\itemsep{0.2em}
    \item \textbf{Gender of artist:} \textit{M} (Man), \textit{W} (Woman), \textit{NB} (Non-binary)
    \item \textbf{Function of artist:} \textit{Feat} (Featured artist), \textit{Main} (Main artist), \textit{Neither}, \textit{Uncredited}
    \item \textbf{Style of vocal delivery:} \textit{R} (Rapped vocals), \textit{S} (Sung vocals), \textit{Spoken}
    \item \textbf{Minimum pitch value:} In Hz
    \item \textbf{First quartile pitch value:} In Hz
    \item \textbf{Median pitch value:} In Hz
    \item \textbf{Third quartile pitch value:} In Hz
    \item \textbf{Maximum pitch value:} In Hz
    \item \textbf{Environment value:} On a scale of E1 to E5
    \item \textbf{Layering value:} On a scale of L1 to L5
    \item \textbf{Width (panning) value:} On a scale of W1 to W5
\end{enumerate}

The annotators determined the name of the artist(s) performing in each section by ear, and using song lyric website Genius.com to validate their hearing. In cases where an artist only provides minimal background vocals (a few words) in a particular formal section, their name is not included. One annotator then provided analytical data on each formal section performed by one artist only. Data on gender was gathered from media interviews and social media statements from the artists, and matches the artist's gender identity at the time of the dataset creation. This methodology yielded three categories: man, non-binary, and woman. We understand these labels as umbrella terms that encompass a variety of lived experiences that intersect with race, sexuality, and other power structures. The style of vocal delivery was determined by ear. The distinction between rapping and singing is porous, with many vocalists adopting ambiguous modes of vocal delivery. We consider any formal section containing a melodic line performed with sustained pitches as sung. 

The pitch data was obtained by first isolating the vocals from the full mix using Open-Unmix\cite{stoter19} and then running the pYIN Smoothed Pitch Track transform \cite{Matthias2014} on the isolated vocal file. The minimum, first quartile, median, third quartile, and maximum pitch points in each formal section were calculated and recorded in the dataset.\footnote{The accuracy of the F\textsubscript{0} estimates used to calculate this feature is impacted by the quality of the vocal source separation. A more accurate isolated vocal file would allow for more precise pitch data. Additionally, since pYIN Smoothed Pitch Track can only track a single melodic line, the accuracy of the pitch data is lessened in sections that feature multiple vocal layers with different pitch content.} 

The Environment, Layering, and Width values were determined by the first annotator to ensure consistency. Rather than attempting to reconstruct the mixing process itself, the annotations for these parameters represent the way a listener might perceive the final mix upon listening to it on stereo speakers. The Environment of a voice is the space in which the voice reverberates. Environment values were determined via an aural analysis of the full track by using the following scale\footnote{The scales were initially published in \cite{Duguay2022}.}:

\setdefaultleftmargin{0.7cm}{}{}{}{}{}
\begin{enumerate}
 \setlength\itemsep{0.5em}
    \item[E1:] The voice’s environment sounds flat. There might be minimal ambiance added to the voice, but there is no audible echo or reverberation.
    \item[E2:] The last word or syllable of most musical phrases is repeated through an echo or reverberation effect.
    \item[E3:] The vocal line is repeated in one clear layer of echo. This added layer may be dry or slightly reverberant and has a lower amplitude than the main voice.
    \item[E4:] The main voice is accompanied by a noticeable amount of reverberation. There is no clear echo layer, but rather a sense that the main voice is being reverberated across a large space.
    \item[E5:] The main voice is accompanied by two or more layers of echo. The echo layers may be noticeably reverberant, similar in amplitude to the main voice, and difficult to differentiate from one another.
\end{enumerate}

The Layering of a voice refers to the additional vocal tracks that are dubbed over a single voice. Layering values were determined via an aural analysis of the full track by using the following scale: 

\setdefaultleftmargin{0.7cm}{}{}{}{}{}
\begin{enumerate}
 \setlength\itemsep{0.5em}
    \item[L1:] The voice is presented as solo. Occasionally, a few words may be doubled with another vocal track for emphasis. Double-tracking is often used in the mixing process to create a fuller sound, with a final result sounding like a single vocal layer. Such cases fall into this category.
    \item[L2:] The voice is presented as solo, but additional vocal layers are added at the end of musical phrases for emphasis.
    \item[L3:] The main voice is accompanied by one or two layers. Layers might provide minimal harmonies or double the main voice. The layers have a noticeably lower amplitude than the main voice.
    \item[L4:] The main voice is accompanied by two or more layers. These layers are close copies of the main voice, sharing the same pitch and similar amplitude.
    \item[L5:] The main voice is accompanied by two or more layers. These layers add harmonies to the main voice, creating a thick and multi-voiced texture.
\end{enumerate}

The Width of a voice refers to the breadth it occupies on the stereo stage. The Width was analyzed aurally with the aid of panning visualisation tool MarPanning\cite{McNally2009}. The annotator simultaneously listened to the isolated vocal audio and observed the MarPanning visualization generated from the isolated vocals to determine the Width value. Since Open-Unmix occasionally omits reverberated components of the voice from the isolated file, the analyst then listened to the full track to confirm the Width value. Width values were determined according to the following scale: 

\setdefaultleftmargin{0.7cm}{}{}{}{}{}
\begin{enumerate}
 \setlength\itemsep{0.5em}
    \item[W1:] The voice occupies a narrow position in the center of the stereo stage.
    \item[W2:] The voice occupies a slightly more diffuse position in the center of the stereo stage.
    \item[W3:] The main voice occupies a narrow position in the center of the stereo stage, but some of its components (echo, reverberation, and/or additional vocal tracks) are panned toward the sides. These wider components have a lower amplitude than the main voice.
    \item[W4:] The main voice occupies a slightly more diffuse position in the center of the stereo stage, and some of its components (echo, reverberation, and/or additional vocal tracks) are panned toward the sides. These wider components have a lower amplitude than the main voice.
    \item[W5:] The main voice and its associated components (echo, reverberation, and/or additional vocal tracks) are panned across the stereo stage. All components have a similar amplitude.
\end{enumerate}

\subsection{Metadata}

The following metadata is provided for each song in the dataset:
\setdefaultleftmargin{0.4cm}{}{}{}{}{}
\begin{enumerate}
 \setlength\itemsep{0em}
    \item \textbf{Index number:} From 1 to 331
    \item \textbf{Year of first appearance on \textit{Billboard} “Hot 100” year-end charts}
    \item \textbf{Chart position:} As it appears on the \textit{Billboard} “Hot 100” year-end charts
    \item \textbf{Song title:} As it appears on the \textit{Billboard} “Hot 100” year-end charts
    \item \textbf{Name of artists:} As it appears on the \textit{Billboard} “Hot 100” year-end charts
    \item \textbf{Collaboration type:}
      \begin{itemize}
       \setlength\itemsep{0em}
      \item \textit{Lead/featured:} Collab. with lead artist(s) and featured artist(s)
      \item \textit{No lead/featured:} Collab. with no determined lead
      \item \textit{DJ/vocals:} Collab. between a DJ and vocalist(s)
      \end{itemize}
    \item \textbf{Gender of artists:}
       \begin{itemize}
        \setlength\itemsep{0em}
      \item \textit{Men:} Collab. between two or more men
      \item \textit{Women:} Collab. between two or more women
      \item \textit{Mixed:} Collab. between two or more artists of different genders
      \end{itemize}
    \item \textbf{Collaboration type + gender:}
        \begin{itemize}
         \setlength\itemsep{0em}
      \item \textit{Collab M:} Collab. between men, no determined lead
      \item \textit{Collab M and W:} Collab. between men and women, no determined lead
      \item \textit{Collab NB and W:} Collab. betwen women and non-binary artists, no determined lead
      \item \textit{Collab W:} Collab. between women, no determined lead
      \item \textit{DJ with M:} Collab. between male DJ and male vocalist
      \item \textit{DJ with Mix:} Collab. between male DJ and mixed-gender vocalists
      \item \textit{DJ with NB:} Collab. between male DJ and non-binary vocalist
      \item \textit{DJ with W:} Collab. between male DJ and female vocalist
      \item \textit{M ft. M:} Men featuring men
      \item \textit{M ft. W:} Men featuring non-binary artist(s)
      \item \textit{W ft. M:} Women featuring men
      \item \textit{W ft. W:} Women featuring women
      \end{itemize}
    \item \textbf{MusicBrainz URL:} Link to the song on open music encyclopedia MusicBrainz
\end{enumerate}

Each song in the dataset is labeled with an index number from 1 to 331. Songs are numbered in reverse chronological order, beginning with the 2019 charts and ending with 2010. One annotator obtained the metadata on year, chart position, title, and artists from the information available on the \textit{Billboard} charts. Within years, songs are organized according to their position on the chart, from highest to lowest. Some songs appear on the charts two years in a row. In such cases, we only include the data for the earliest appearance. 

\begin{figure}[t]
 \includegraphics[width=\columnwidth]{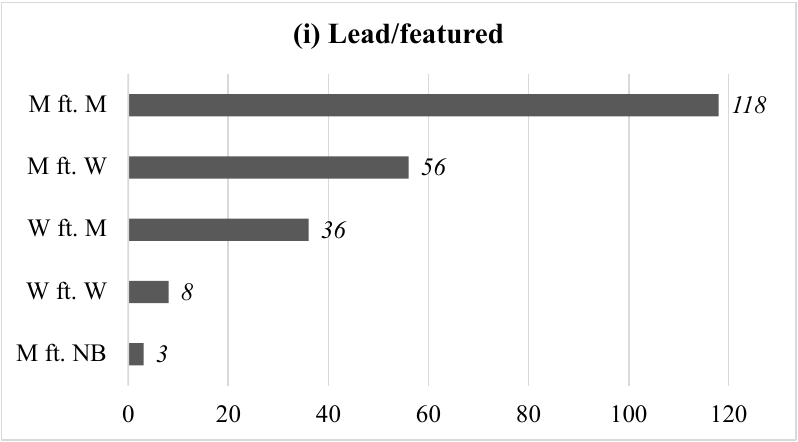}\hfill
 \includegraphics[width=\columnwidth]{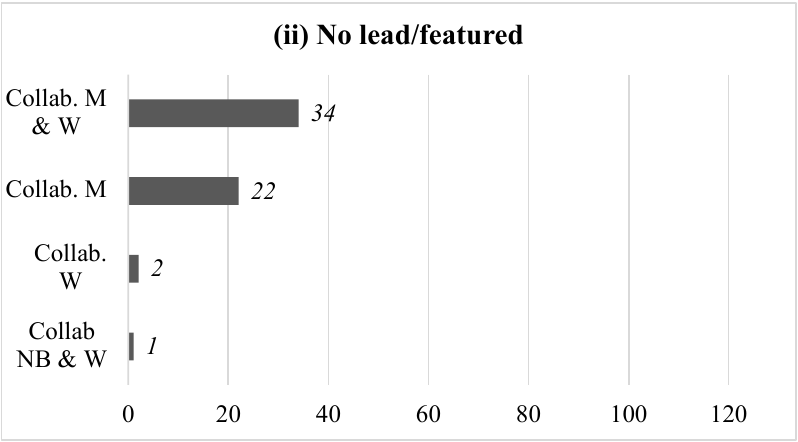}\hfill 
 \includegraphics[width=\columnwidth]{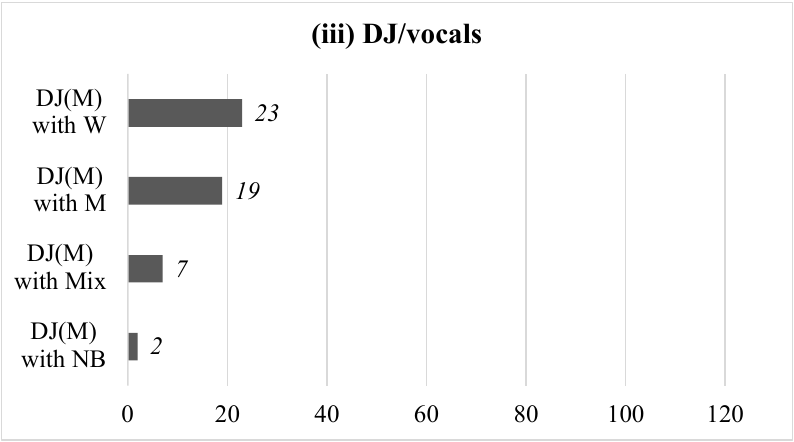}\hfill  
 \caption{Summary of the gender distribution across different types of multi-artist collaborations. Subplot (i) shows gender counts for collaborations with lead and featured artists, subplot (ii) shows collaborations with no determined lead or featured artist, and subplot (iii) shows collaborations between DJs and vocalist(s).}
 \label{fig:statistics}
\end{figure}

\subsection{Corpus Statistics}

The dataset can be divided into three categories (shown in Figure \ref{fig:statistics}): (i) collaborations between the lead artist(s) and featured artist(s), which account for 221, or 66.7\% of the tracks, (ii) collaborations with no determined lead or featured artist, which account for 59, or 17.8\%, of the tracks, and (iii) collaborations between a DJ and a vocalist, which account for 51, or 15.4\% of the tracks. In category (i), the lead artist usually performs the majority of vocals. For example, in “No Limit” (2018) by G-Eazy ft. A\$AP Rocky \&  Cardi B, G-Eazy performs most of the vocals. A\$AP Rocky accompanies him in the chorus and Cardi B raps the second verse. In category (ii), the performance of the vocals is often more equally distributed. Such collaborations are often billed as “duets,” and the artists’ names are separated by a ``+'', a ``\&'', or a comma on the \textit{Billboard} charts. For example,``Something’ Bad'' (2014) is labeled as a ``Miranda Lambert Duet With Carrie Underwood.'' Both vocalists perform approximately equal portions of the song. In category (iii), the DJ does not provide vocals. In ``Sweet Nothing'' (2012), for instance, only the featured Florence Welch sings. The voice of DJ Calvin Harris is not heard. 

Mixed-gender collaborations (including any combination of non-binary, women, and men artists) frequently appear on the Billboard charts and account for 162, or 49\%, of the tracks in the dataset. Collaborations between two or more men account for 159 tracks, or 48\% of the dataset. Finally, collaborations between women account for 10, or 3\%, of the tracks. In six of the ten years under study--2011, 2012, 2015, 2017, 2018, and 2019--no collaborations between women reached the \textit{Billboard} “Hot 100” year-end chart. Conversely, songs with two or more male vocalists were a consistent fixture on the charts. Mixed-gender collaborations, with any combination of men, women, and non-binary artists within the same track, also frequently appear on the charts. 

\begin{figure}[t]
 \centerline{
 \includegraphics[width=\columnwidth]{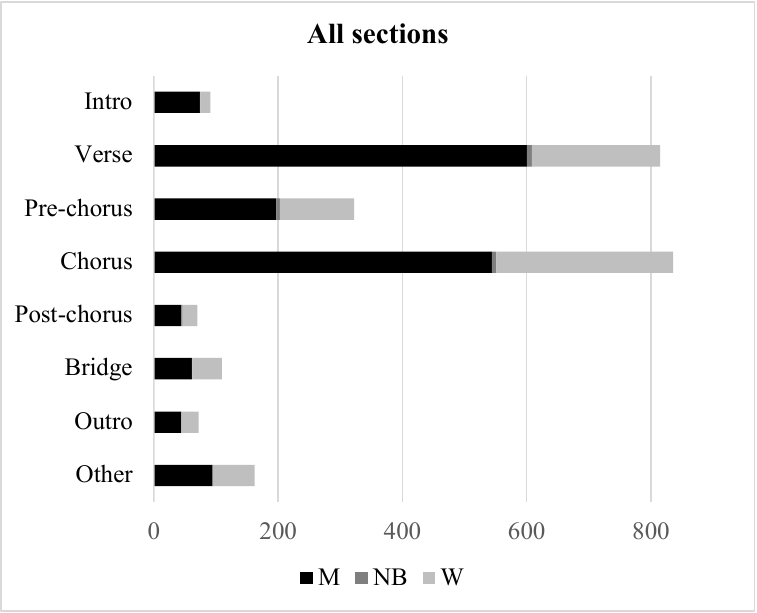}}
 \caption{Number of formal sections performed by a single artist (main, featured, or neither), categorized according to formal section type}
 \label{fig:all}
\end{figure}

\begin{figure}[t]
 \centerline{
 \includegraphics[width=\columnwidth]{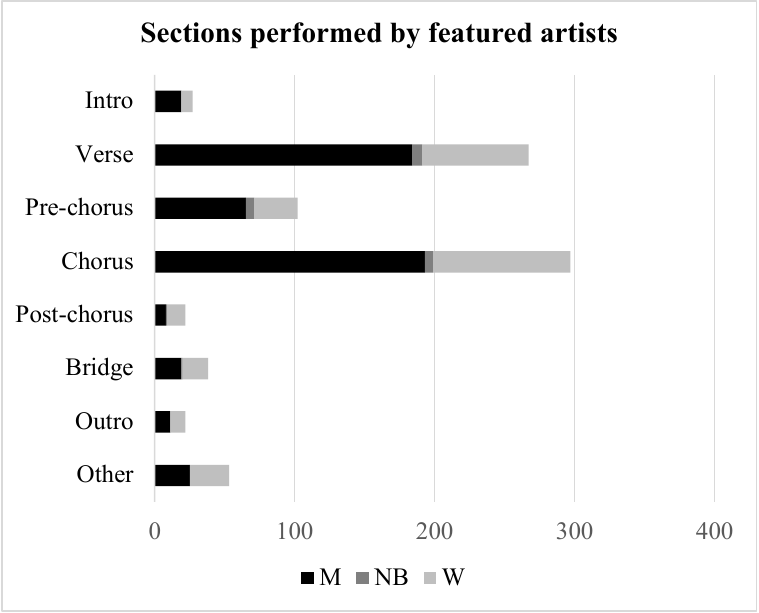}}
 \caption{Number of formal sections performed by featured artists, categorized according to formal section type}
 \label{fig:feat}
\end{figure}

Figure \ref{fig:all} shows the number and type of sections performed by individual artists in the corpus, categorized according to gender. This figure includes identical sections (such as choruses) that are repeated within a song. Sections in which more than one artist performs are not included. More sections are performed by men than by women and non-binary artists, which is to be expected given the over-representation of men in the dataset as a whole (Figure \ref{fig:statistics}). Figure \ref{fig:feat} displays the number and type of sections performed by \textit{featured} artists only.

\section{Experiment: Vocal Production Features and Gender}

This section examines the relationship between the gender of an artist and the treatment of their voice, as characterized by three of the annotated musical features in the dataset: Environment, Layering, and Width. For the purposes of statistical power in the experiment, only songs with men and/or women artists were included. We only included tracks that contained verse and chorus sections to remove section types that occur in only a few tracks. In order to avoid over-representations of tracks with repeated sections (i.e., several instances of the same chorus), we sampled the first verse and chorus performed by a single artist from each track.\footnote{If the first verse of a song was performed by two artists simultaneously, while the second verse was only performed by one, we sampled the second verse.}This method resulted in the inclusion of two sections from {287 of the 331 dataset tracks in the experiment. 

We analyzed the data with three separate logistic regressions--one for each feature--using the statsmodels package in Python. We encoded the different levels of the parameter scales (defined in Section 3.1) with one-hot encoding in order to allow us to examine whether there is a correspondence between specific parameter scale levels and gender.

Of the three logistic regressions, Environment (\emph{R\textsuperscript{2}}\textsubscript{McFadden} (4, \emph{N} = 574) = 0.028, \emph{p} < 0.0001) and Width (\emph{R\textsuperscript{2}}\textsubscript{McFadden} (4, \emph{N} = 574) = 0.035, \emph{p} < 0.0001) were statistically significant, while Layering (\emph{R\textsuperscript{2}}\textsubscript{McFadden} (4, \emph{N} = 574) = 0.0036, \emph{p} = 0.64) was not. The McFadden \emph{R\textsuperscript{2}} values for both Environment and Width were very low. This was not surprising since we did not anticipate that these features, particularly in isolation, would be explanatory. We were instead interested in exploring whether there is a significance between these features with respect to the man/woman gender binary in these collaborations.

For Environment, there were significant effects (\emph{p} < 0.0001) for E1 (\textbeta=-1.18, 95\%CI [-1.49, -0.87]), E2 (\textbeta=-1.12, 95\%CI [-1.56, -0.69]), and E3 (\textbeta=--0.78, 95\%CI [-1.14, -0.42). There was a significant negative effect for the lower/mid-level environment values and gender, meaning that men's voices are more likely to be set in less reverberant spaces than women's voices. For Width, there were significant effects at all of the levels: W1 (\textbeta=-1.84, 95\%CI [-2.50, -1.17]), W2 (\textbeta=-1.58, 95\%CI [-2.39, -0.77]), W3 (\textbeta=--1.13, 95\%CI [-1.51, -0.75]), W4 (\textbeta=-0.47, 95\%CI [-0.77, -0.17), and W5 (\textbeta=-0.60, 95\%CI [-0.95, -0.25]). 

The Width results are harder to interpret than the Environment ones because the coefficient values are smaller and all negative. This is likely due to the imbalance between men and women in featured artist roles, both in the dataset (see Figure \ref{fig:statistics}) overall and in the sample used in this experiment (404 of the included sections featured men while only of 170 featured women). However, the overall trend is similar to the one in the Environment experiment: lower-level values are more common for men than women. Men's voices are more likely to occupy a narrow, centered position on the stereo stage, while women's voices are more likely to occupy a wider space. These results were expected given that high Environment values tend to be associated with high Width values, as the reverberated components of a voice are generally panned across the stereo stage.  

The lack of significant results for Layering indicates that there are no differences in the ways in which this parameter is applied to men's and women's voices. Since textural variation (such as the addition of vocal layers) is a standard feature of verse-chorus form, it is possible that Layering is linked to the type of formal section rather than to the gender of the vocalist. The significant results for the Environment and Width parameters can be interpreted in light of Br{\o}vig-Hanssen's and Danielsen's work on technological mediation\cite{BH2016}. The authors establish a distinction between transparent and opaque technological mediation in recorded music. Transparent mediation, on one hand, is meant to create a recorded product that sounds natural and unaltered. Low Environment and Width values, for instance, are closer to transparent mediation because they sound closer to a real-life performance that is unmediated with artificial reverb or panning. Opaque mediation, on the other hand, highlights the use of technology by making it obvious to the listener. High Width and Environment values, with their clearly audible artificial reverberation and wide panning, are examples of opaque mediation. The results of the experiment therefore suggest that men’s voices are more likely to be mixed to sound “transparent” and natural while women’s voices are more likely to be mixed to sound “opaque” and technologically mediated.  

 Overall, this experiment demonstrates that within verse and chorus sections in CoSoD, there is a significant difference between the treatment of men's and women's vocals in terms of Environment and Width. This suggests that some mixing parameters contribute to the sonic differentiation of men’s and women’s voices in popular music.

\section{Conclusion}
CoSoD is a 331-song corpus of all multi-artist collaborations for faciliating appearing on the 2010--2019 \textit{Billboard} “Hot 100” charts. Each song in the dataset is annotated with metadata,  formal sections, and aspects of vocal production (including reverberation, layering, panning, and gender of the artists). As outlined in Section 2, CoSoD has several implications for MIR research. It provides annotated data for structural segmentation tasks and a listener-centered perspective on vocal mixing that could be useful for automatic music mixing tasks. The dataset could also be used to determine how these parameters interact with song form. Further study could also examine the relationship between the vocal range of an artist in a given section, their type of vocal delivery (rapped, spoken, or sung), and mixing parameters. Finally, the dataset also allows for the examination of the ways in which Environment, Layering, and Width values tend to be grouped together to create specific vocal production effects.


The dataset also facilitates musicological study of multi-artist collaborations post-2010 and gender norms. The experiment in Section 4 demonstrates this, as its results suggest that, for the chorus and verse data sampled from 287 songs in the dataset, men's voices are more likely to be narrow and less reverberated than women's. Opportunities for future research include examining whether there is a significant difference in the way Environment, Width, Layering, or other parameters are applied to women's and men's voices \textit{within} collaborations that feature mixed- and same-gender vocalists. In other future work, we plan on expanding the annotations in the dataset with time-aligned lyrics, harmonic analyses, and additional performance data for the voice extracted using AMPACT \cite{devaney2012, devaney2016}. These annotations will include both spectral features and semantic descriptors, and the data will be encoded in relation to vocal-line transcriptions, where possible \cite{devaney2020}. We also plan on providing annotations on vocal production parameters in sections performed by multiple artists and examining how vocal production parameters correlate with mixing parameters such as panning.

Finally, while our dataset focuses on gender, we are also interested in encoding other aspects of identity, such as race, in order to provide an intersectional perspective on artists' identities. However, categorizing artists according to race proves to be more problematic than gender. Matthew D. Morrison writes that “white (and other nonblack) people freely express themselves through the consumption and performance of commodified black aesthetics without carrying the burden of being black under white supremacist structures” \cite[p. 791]{Morrison2019}. In other words, white and non-Black artists--such as rappers Iggy Azalea and G-Eazy, or singer Bruno Mars--often assume particular sonic characteristics that implicitly associate them with commodified notion of Blackness. By categorizing all white artists together, for instance, we would ignore this phenomenon and the way it is sonically realized.  Further work needs to be done to understand how to best expand on CoSoD, or datasets in general, to account for this dynamic. 

\bibliography{ISMIRtemplate}

%
%
%
%
%

\end{document}